\documentclass[prb,twocolumn,showpacs,preprintnumbers,amsmath,amssymb, superscriptaddress]{revtex4}

\usepackage{graphicx}
\usepackage{graphics}
\usepackage{dcolumn}
\usepackage{bm}
\usepackage{subfigure}

\begin{document}


\title{De Haas-van Alphen oscillations in the charge-density wave compound lanthanum tritelluride (LaTe$_3$)}

\author{N. Ru}
\affiliation{Geballe Laboratory for Advanced Materials and Dept. of Applied Physics, Stanford University, Stanford, CA 94305 (USA)\\
}

\author{R. A. Borzi}
\altaffiliation[Present address: ]{INIFTA (UNLP-CONICET), c.c. 16, Suc. 4, 1900 La Plata, Argentina; and
Departamento de Fisica, IFLP, UNLP, c.c. 67, 1900 La Plata, Argentina.}
\author{A. Rost}
\author{A. P. Mackenzie}
    \affiliation{School of Physics and Astronomy, University of St. Andrews, St. Andrews KY16 9SS (UK)}

\author{J. Laverock}

\author{S. B. Dugdale}
\affiliation{H. H. Wills Physics Laboratory, University of Bristol, Tyndall Avenue, Bristol BS8 1TL (UK)}

\author{I. R. Fisher}
    \affiliation{Geballe Laboratory for Advanced Materials and Dept. of Applied Physics, Stanford University, Stanford, CA 94305 (USA)\\
}

\date{\today}

\begin{abstract}

De Haas-van Alphen oscillations were measured in lanthanum tritelluride (LaTe$_3$) to probe the partially gapped Fermi surface resulting from charge density wave (CDW) formation.  Three distinct frequencies were observed, one of which can be correlated with a FS sheet that is unaltered by CDW formation.  The other two frequencies arise from FS sheets that have been reconstructed in the CDW state.
\end{abstract}


\maketitle

\section{Introduction}
In materials with Fermi surfaces that are not purely one-dimensional, the formation of a charge density wave below a certain transition temperature will often create a ``partially gapped'' Fermi surface by nesting only certain regions of the Fermi surface---those that are best spanned by the charge density wavevector.  The material remains metallic below the transition, with transport properties that are altered by an often significantly changed Fermi surface.

\begin{figure}
\includegraphics[width = 3.1 in]{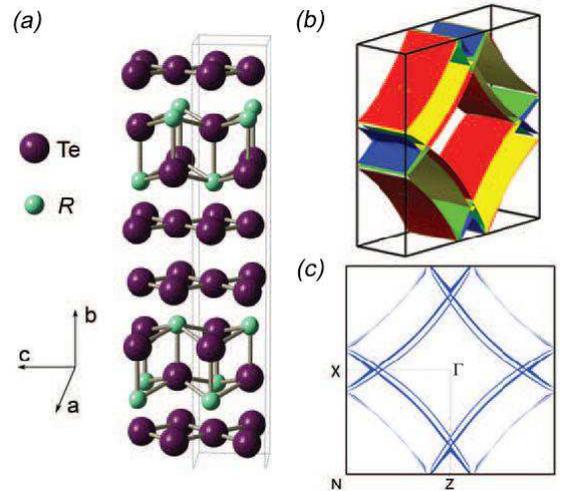}
\caption{\label{fig:multifigure}(color online) (a)The crystal structure of $R$Te$_3$. (b) The Fermi surface of LaTe$_3$ as calculated by LMTO. (c) Projection down the $b^*$-axis for the LaTe$_3$ Fermi surface. Symmetry points for the $\Gamma$-X-Z plane ($k_y$ = 0) are labeled. Dispersion of the FS along $b^*$ is indicated by the thickness of the lines.}
\end{figure}

The rare-earth tritellurides have recently emerged as a prototypical family of CDW compounds.\cite{dimasi_1995, malliakas_2006, ru_CDWtransition, brouet_long} In the unmodulated state, the materials have a crystal structure (Fig.~\ref{fig:multifigure}(a)) that is layered and weakly orthorhombic (space group Cmcm)\cite{norling}, consisting of double layers of nominally square-planar Te sheets, separated by corrugated $R$Te slabs.   In this space group, the long $b$-axis is perpendicular to the Te planes.  The band structure has been calculated for the unmodulated structure, yielding a simple FS (Fig.~\ref{fig:multifigure}(b) and (c)) consisting of slightly warped inner and outer diamond sheets formed from $p_x$ and $p_z$ orbitals of Te atoms in the square planar layers, both doubled due to the effects of bilayer splitting, and with minimal $b$-axis dispersion.\cite{laverock}  The susceptibility $\chi(q)$ has peaks of similar magnitude in both in-plane directions for $q$ $\approx$ 2/7 $c^*$ and  2/7 $a^*$ ($c^*$ = 2$\pi/c$).\cite{yao, johannes}

All compounds in the $R$Te$_3$ family exhibit a simple unidirectional lattice modulation with a wavevector $q_1$ $\approx$ 2/7 $c^*$, corresponding to the maximum in $\chi(q)$.\cite{dimasi_1995, malliakas_2006, ru_CDWtransition, brouet_long, brouet, iyeiri, kim, fang}  Temperatures for this transition range from $T_{c1}$ = 244 K for TmTe$_3$ to $T_{c1}$ = 416 K for SmTe$_3$, with those for the as yet unmeasured $R$ = La, Ce, Pr, Nd presumed to be even higher.  Angle resolved photoemission spectroscopy (ARPES) has revealed that portions of the FS which are nested by these wave vectors are indeed gapped, implying that the gain in one-electron energy contributes to CDW formation.\cite{gweon, brouet, komoda}  Below the transition, the Fermi surface is partially gapped and exhibits metallic behavior.\cite{iyeiri, ru_2006, ru_CDWtransition}  For the heaviest rare earths($R$ =  Dy, Ho, Er, Tm), a second transition is observed, which has been shown for ErTe$_3$\cite{ru_CDWtransition} and HoTe$_3$\cite{Ho_xray} to have a wavevector $q_2$ $\approx$ 1/3 $a^*$. This second CDW wavevector gaps additional portions of the Fermi surface,\cite{moore_comm} with electronic behavior remaining metallic below both transitions. \cite{ru_CDWtransition}

In this paper we present results of experiments probing the FS of LaTe$_3$ in the CDW state, and compare these results with band structure calculations for the unmodulated structure.  LaTe$_3$ exhibits just one CDW state with wavevector $q_1$ $\approx$ 2/7 $c^*$ with a transition temperature over 400 K.  Its reconstructed Fermi surface is expected to be similar to that of CeTe$_3$, which has been investigated by ARPES by V. Brouet and coworkers.\cite{brouet}  Although these measurements reveal the presence of CDW shadow bands in the CDW state, the precise shape and area of the reconstructed FS remains to be determined.    De Haas-van Alphen (dHvA) oscillations in magnetization, which provide a measure of the area enclosed by the reconstructed FS, are uniquely suited for this purpose.

\section{Experimental Methods}

Single crystals of LaTe$_3$ were grown by slow cooling a binary melt as described previously.\cite{ru_2006}  Residual resistivity ratios (RRR = $\rho$(300 K)/$\rho$(1.8 K)) of up to 120 were observed, which is an indicator of good crystal quality.

AC susceptibility measurements were made using two counterwound pickup
coils consisting of approximately 1000 turns of 12 $\mu$m-diameter
copper wire. The primary is a superconducting coil that is part of the
main magnet; it was driven at a frequency near 83 Hz, providing an
AC field of 3.3$\times 10^{-5}$ T rms. A LaTe$_3$ crystal of mass 0.7 mg was
inserted in one of the secondary coils with the $b$-axis parallel to
the field. The sample was held in place with grease, and thermally
grounded with gold and copper wires to the mixing chamber of the
dilution refrigerator (Oxford Instruments Kelvinox 25).
Low-temperature transformers mounted on the 1-K pot were used to
provide an initial signal boost of approximately a factor of 100.
Measurements were performed at a temperature of 40 mK. The magnetic
field was swept from 15 T to 0.5 T at a rate of 0.025 T/min, with a
data density of one point per 0.6 Oe.

Cantilever torque measurements were made in a 14 T Quantum Design Physical Properties Measurement System (PPMS) with a base temperature of 1.9 K.  Crystals were shaped to square platelets weighing 1-2 mg.  The magnetic field was set to sweep from 14 to 2 T at a rate of 0.063 T/min, with a data density of one point per 75 Oe.


These torque measurements were repeated with the crystal rotating about an in-plane axis.  Orientation by x-ray diffraction with a Panalytical X'Pert Diffractometer in reflection geometry revealed that the $a$-axis was the axis of rotation i.e. that the field was oriented in the $b$-$c$ plane of the crystal. However, although the crystal used for this study was a well-formed faceted crystal, LaTe$_3$ is prone to twinning due to the weak Van der Waals bonding between the double Te sheets.   Given that the penetration depth of the x-rays is only on the order of microns, it is possible that the sample also contained undetected grains for which the $c$-axis was the axis of rotation.

\section{Results}
\begin{figure}
\includegraphics[width=3in]{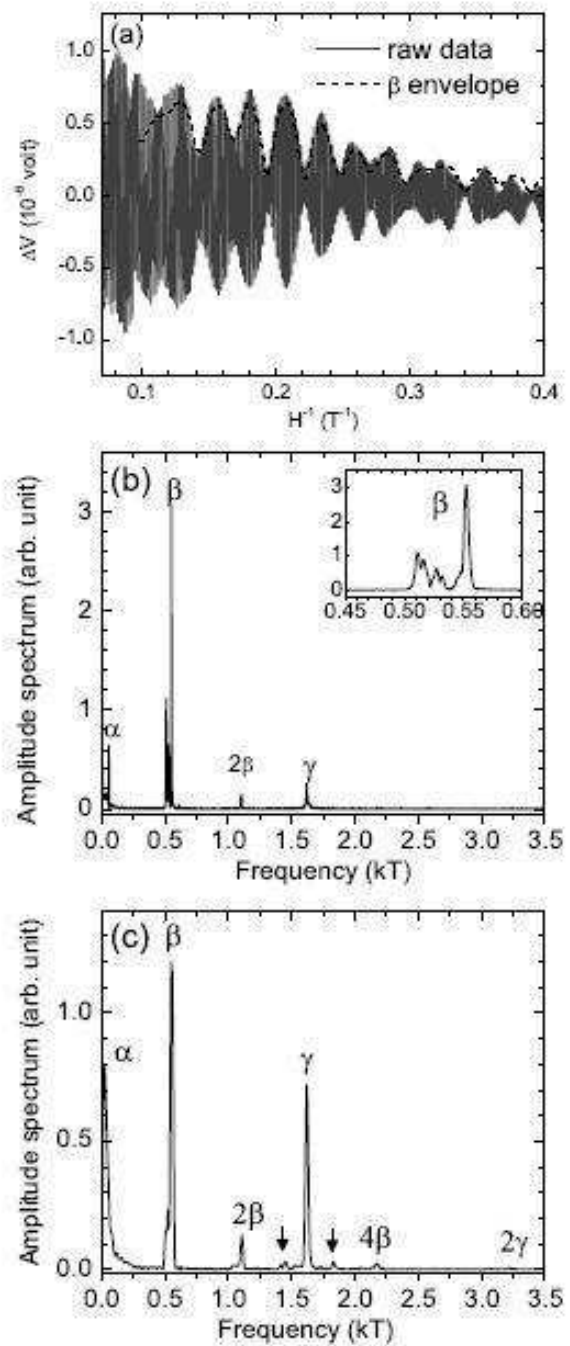}
\caption{\label{fig:dilution}(a) The susceptibility of LaTe$_3$ at T = 40 mK showing dHvA oscillations(gray, solid), with background subtracted. The envelope of the $\beta$ frequency is shown (black, dashed).  (b) Fourier transform over the field range 1.5 to 15T. Inset shows in detail the multiple peaks at $\beta$.  (c)  Fourier transform over the field range 7 to 15T. The principal frequencies $\alpha$, $\beta$, and $\gamma$ and higher harmonics are labeled.  Arrows indicate additional frequencies as described in the text.}
\end{figure}

Clear oscillations were observed in both the ac susceptibility and torque measurements.  At least three distinct frequencies were observed for fields oriented parallel to the $b$-axis: $\alpha$ = 0.058 kT, $\beta$ = 0.52 kT, and $\gamma$ = 1.6 kT, with the $\beta$ frequency composed of two or more very closely spaced frequencies.

Quantum oscillations as seen in the ac susceptibility measurement at 40 mK are shown in Fig.~\ref{fig:dilution}(a), plotted as a function of inverse field, with a background subtracted.  Oscillations were noticeable in the data for fields greater than 1 T.  The predominant frequency seen in the raw data is the $\beta$ frequency with a beating that arises from its closely split peaks.  To emphasize the contribution of $\beta$ to the total oscillatory content, the envelope of $\beta$, derived from the Fourier transform of the raw data, is overlaid in Fig.~\ref{fig:dilution}(a).

   Fig.~\ref{fig:dilution}(b) shows a Fourier transform of Fig.~\ref{fig:dilution}(a) over the field range  1.5 to 15 T.  A Hamming window was used for the Fast Fourier Transform (FFT).  The frequencies $\alpha$, $\beta$, and $\gamma$ are labeled.  The $\beta$ frequency is larger in amplitude than the others and is composed of a cluster of peaks with frequencies ranging from 511 to 553 T, as shown more clearly in the inset to Fig.~\ref{fig:dilution}(b).  In this Fourier transform, the $\gamma$ peak has some structure, but, unlike $\beta$, is not distinctly split into separate peaks.  Also labeled in the Fourier transform is the second harmonic 2$\beta$.

  A Fourier transform of the data from 7 to 15 T is shown in Fig.~\ref{fig:dilution}(c).  With a shorter field range, the higher frequency components in the data are emphasized.  Higher harmonics 4$\beta$ and 2$\gamma$ of the fundamental frequencies are clearly visible, as well as small peaks at 1.45 and 1.73 T on either side of $\gamma$, which are labeled with arrows.   The shortened field range also reduces the Fourier transform resolution, partially obscuring the $\beta$ peak splittings.

\begin{figure}
\includegraphics[width =3.1 in]{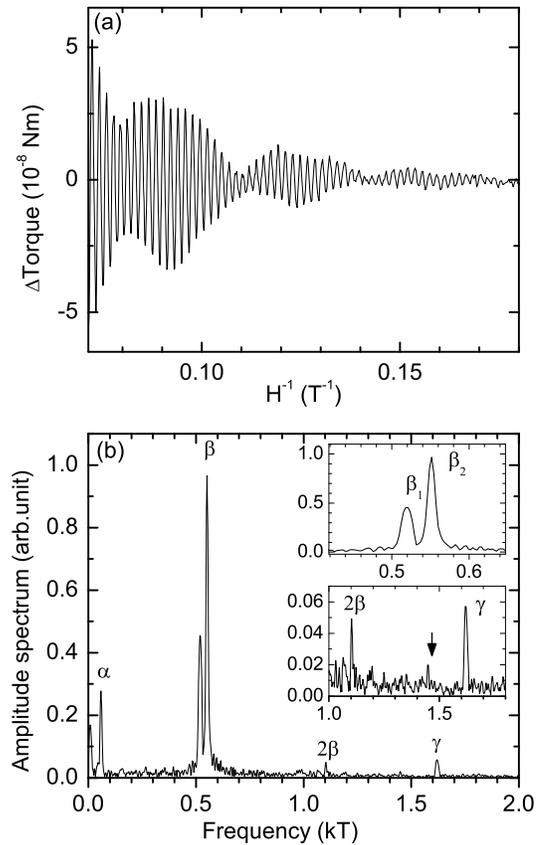}
\caption{\label{fig:PPMS}(a)Torque data for LaTe$_3$ at T = 2 K with field oriented 5$^\circ$ from the $b$-axis, with polynomial background subtracted.  (b) Fourier transform from 4 to 14T.  The principal frequencies $\alpha$, $\beta$, and $\gamma$ and the higher harmonic 2$\beta$ are labeled.  The insets show the double peaks $\beta_1$ and $\beta_2$, as well as a very small peak at 1.45 kT marked by an arrow. }
\end{figure}

In order to make a detailed angular and effective mass study, additional measurements were performed using a cantilever torque magnetometer and a second crystal of LaTe$_3$.  Fig.~\ref{fig:PPMS}(a) shows the results of a torque magnetometer measurement performed at 2.0 K with the $b$-axis of the crystal offset from $H$ by 5$^\circ$ to provide a torque of suitable magnitude.  A third-order polynomial background has been subtracted.  Oscillations were noticeable in the data for fields greater than 4 T.

The Fourier transform of the torque data is shown in Fig.~\ref{fig:PPMS}(b).  Again, the predominant frequencies are labeled as $\alpha$, $\beta$, and $\gamma$.  The spectrum is very similar to that in Fig.~\ref{fig:dilution}, despite the reduced field range and the higher temperature.  Here, the $\beta$ frequency is clearly split into two, labeled in the inset to Fig.~\ref{fig:PPMS}(b) as $\beta_1$ and $\beta_2$.  Given the smaller field range, sparser point density, and smaller signal to noise levels, higher harmonics above 2 kT were not observed.

The frequencies $\alpha$, $\beta$, and $\gamma$ correspond to closed orbits with cross-sectional areas that can be expressed in terms of Brillouin zone (BZ) area, defined by $a^* \times c^* = 2.027 \AA^{-2}$, where $a = 4.4045\AA$ and $c = 4.421\AA$.\cite{Lalattice}  The smallest frequency, $\alpha$, corresponds to 0.28(1)\% of the BZ.  The two peaks $\beta_1$ and $\beta_2$ correspond to 2.44(1)\% and 2.60(1)\% of the BZ, respectively, and the largest frequency $\gamma$ corresponds to 7.65(5)\% of the BZ. These results are summarized in Table~\ref{frequencies}.
\begin{table}
\begin{ruledtabular}
\begin{tabular}{c@{}c|c@{}d@{}d@{}d}
 \multicolumn{2}{c|}{\textbf{AC Susceptibility}} & \multicolumn{4}{c}{\textbf{Torque}} \\
  & F(kT) & & \multicolumn{1}{c}{F(kT)}& \multicolumn{1}{c}{Area(\%BZ)} &\multicolumn{1}{c}{$m^*$ ($m_0$)}\\
\hline
$\alpha$    &0.058(1)     &$\alpha$  &0.059(2)      & 0.28(1)  &0.171(5)\\
$\beta$     &0.511-0.553 &$\beta_1$  &0.519(2)       & 2.44(1)  &0.175(1)\\
 &                     & $\beta_2$    &0.551(2)     & 2.60(1)  &0.183(1)\\
2$\beta$    &1.10(1)     &$2\beta$   &1.11(1)          &2.62(3)^\dag    &0.31(2)\\
$\gamma$    & 1.61-1.62     &$\gamma$  &1.62(1)        & 7.65(5)  &0.39(1)\\
$4\beta$& 2.17(1) &&&&\\
 2$\gamma$ &3.24(5)      &&&&\
\end{tabular}
\end{ruledtabular}
\caption{\label{frequencies} Fundamental dHvA frequencies obtained from susceptibility and torque measurements of LaTe$_3$, corresponding fractional areas of the FS, and the associated cyclotron masses as obtained from Lifshitz-Kosevich fits to the temperature-dependent torque data. $^\dag$The 2$\beta$ frequency being a second harmonic, its FS area is shown as half the value given by the Onsager relation.}
\end{table}

\begin{figure}
\centering
\subfigure{\includegraphics[width = 3.4 in]{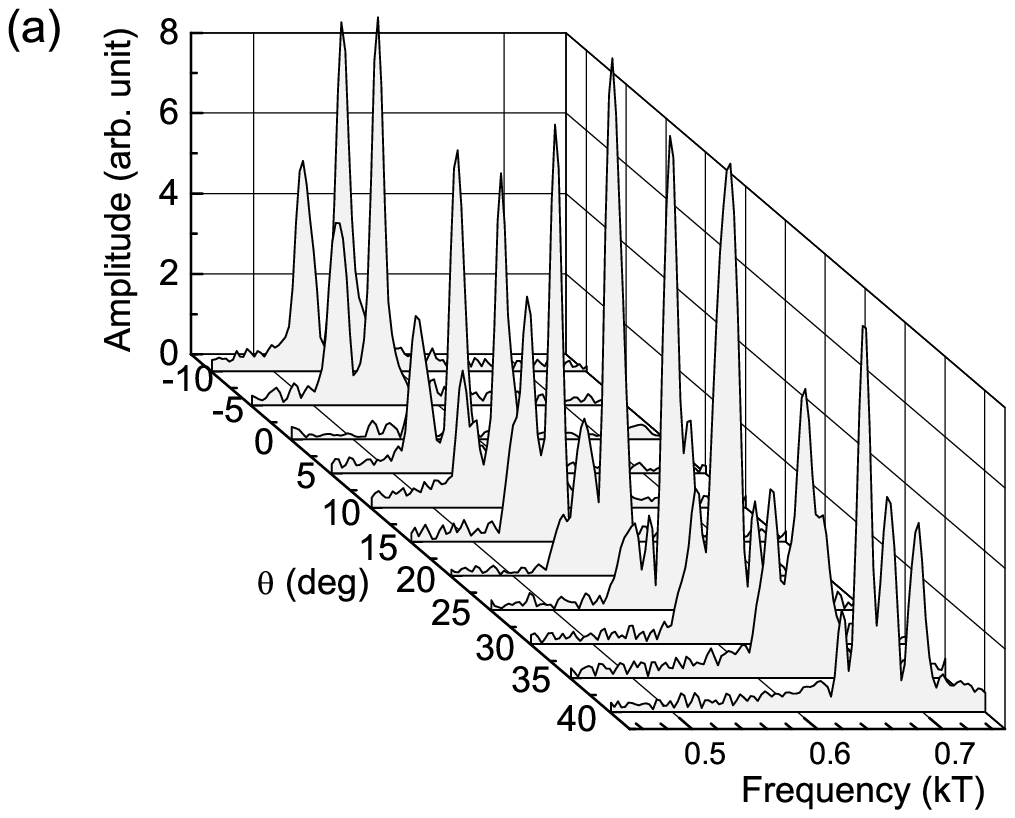}}
\subfigure{\includegraphics[width = 3.3 in]{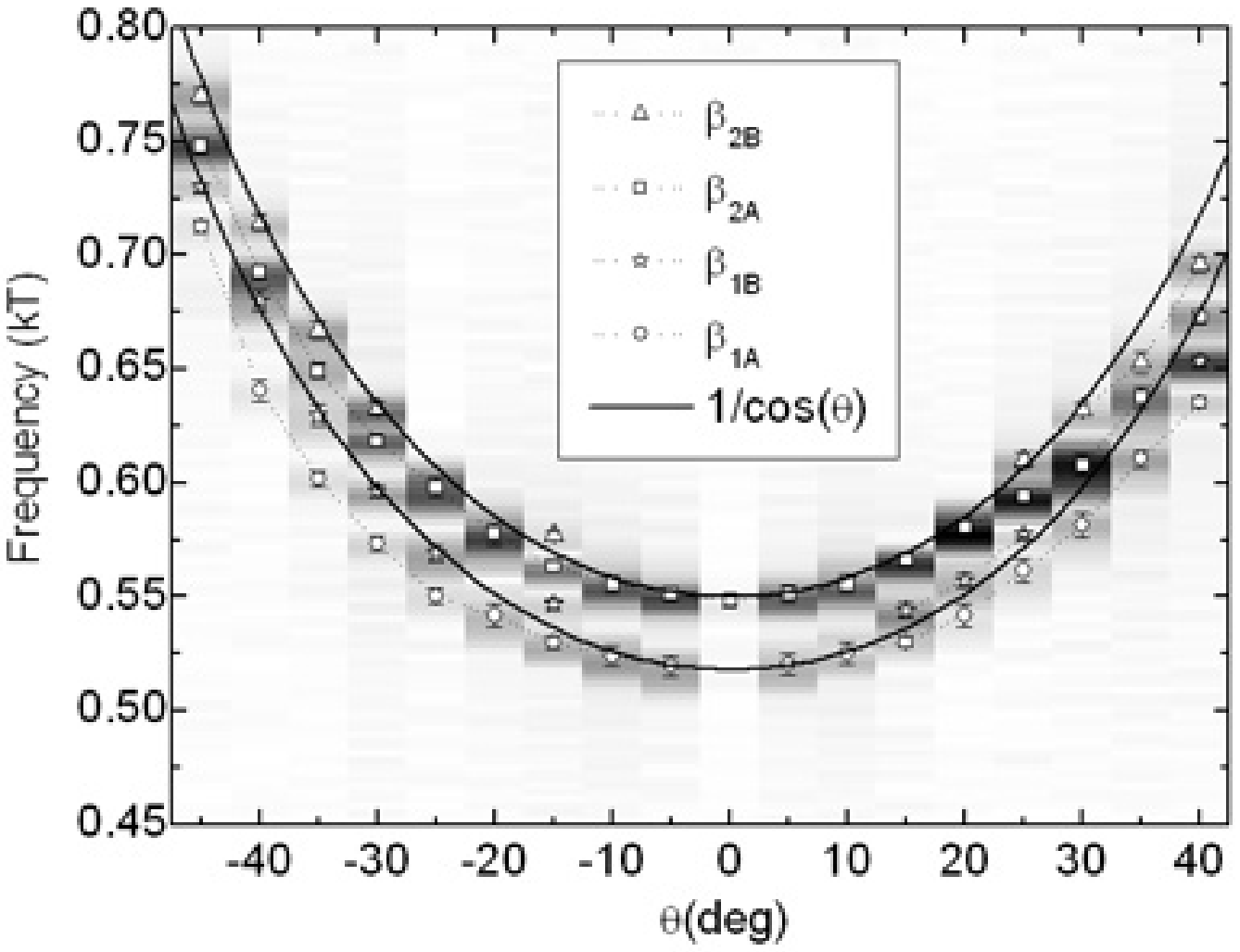}}
\caption{\label{fig:angles_raw}Angle dependence of the $\beta$ frequencies from torque measurements at 2 K. The field is oriented in the $b$-$c$ plane at angles $\theta$ from the $b$-axis. (a)Fourier transforms from $\theta$ = $-10^\circ$ to $40^\circ$, taken over the range 4 to 14 T.  (b)An overhead view of the Fourier transforms from -45$^\circ$ to 40$^\circ$. Fourier transforms are shown as background shadings, with data markers positioned at the center of each peak.  Error bars representing the full width at half max. Solid curves show $1/\cos\theta$.}
\end{figure}

Quantum oscillations in the torque were followed as a function of angle as the field was rotated in the $b$-$c$ plane.  Data were taken at 2 K and at every $5^\circ$ from $\theta$ = $-45^\circ$ to $40^\circ$, where $\theta$ is measured from the $b$-axis.  Fourier transforms were taken over the range 4 to 14 T and the results are shown in Fig.~\ref{fig:angles_raw}(a) for $\theta$  = $-10^\circ$ to $40^\circ$, scaled to focus on the $\beta$ frequencies.  While $\beta$ consists of two peaks at angles near zero, at higher angles it consists of three or four peaks.  For example, in the bottom panel, at 5$^\circ$, $\beta$ consists of two clean peaks.  As the angle is increased to 10$^\circ$, 15$^\circ$, and 20$^\circ$, a shoulder begins to form on the leftmost peak, $\beta_1$, until at 25$^\circ$, $\beta_1$ itself consists of two distinct peaks.  Meanwhile, $\beta_2$ becomes noticeably broader between 15$^\circ$ and 20$^\circ$ until at 25$^\circ$ a peak splits off from the right side of $\beta_2$.  While four peaks are clearly seen at 25$^\circ$, on increasing to 30$^\circ$ the two central peaks have merged, and for larger angles although four peaks are again observed it is difficult to unambiguously identify the origin of each peak.

This information is consolidated in Fig.~\ref{fig:angles_raw}(b) where the Fourier transforms (over the range 4 to 14 T) are shown as a background shading with darker regions representing higher intensities.  Overlaid are markers representing peak positions; the full-width at half-maximum (FWHM) of each peak is represented by the total span of the error bars.  An effort has been made to identify which peaks merge or split into other peaks as the angles change.  As angles increase or decrease from zero, $\beta_1$ splits into two peaks $\beta_{1A}$ and $\beta_{1B}$, while $\beta_2$ splits into two peaks $\beta_{2A}$ and $\beta_{2B}$.  For comparison, the function $1/\cos\theta$, which represents the angle dependence of a FS sheet with no $b$-axis dispersion, is shown as solid curves. The close correspondence to $1/\cos\theta$ indicates that the FS sheets are indeed quasi-2D with minimal $b$-axis dispersion, as indicated by band structure calculations.  However, there are small deviations from a $1/\cos\theta$ dependence:  $\beta_{1B}$ and $\beta_{2B}$ follow $1/\cos\theta$ very well, while $\beta_{1A}$ and $\beta_{2A}$ are ``shallower'' than $1/\cos\theta$.  This implies that the latter two frequencies result from FS maxima around which the curvature is slightly larger than in other parts of the FS.

\begin{figure}
\centering
\includegraphics[width = 3.3 in]{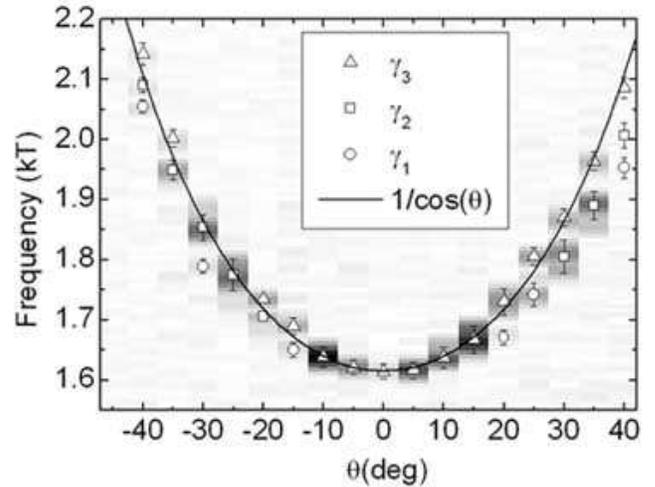}
\caption{\label{fig:image1600T} Angle dependence of the $\gamma$ frequencies from torque measurements at 2 K. The field is oriented in the $b$-$c$ plane at angles $\theta$ from the $b$-axis.  Fourier transforms for each angle are marked by background shadings. Data markers are positioned at the center of each peak, with error bars representing the FWHM.  Solid curves show $1/\cos\theta$.}
\end{figure}

Fig.~\ref{fig:image1600T} shows a similar plot for the $\gamma$ frequency for Fourier transforms taken over the range 7 to 14 T.  While $\gamma$ is a single peak close to $\theta$ = 0, it soon splits into two and three peaks at higher angles.  An attempt was made to identify $\gamma_1$, $\gamma_2$, and $\gamma_3$, as they split and merge with increasing $\theta$.  The frequency $\gamma_3$ seems to follow the $1/\cos\theta$ dependence very closely, while $\gamma_1$ and $\gamma_2$ are shallower than $1/\cos\theta$.

\begin{figure}
\centering
\includegraphics[width=3.3in]{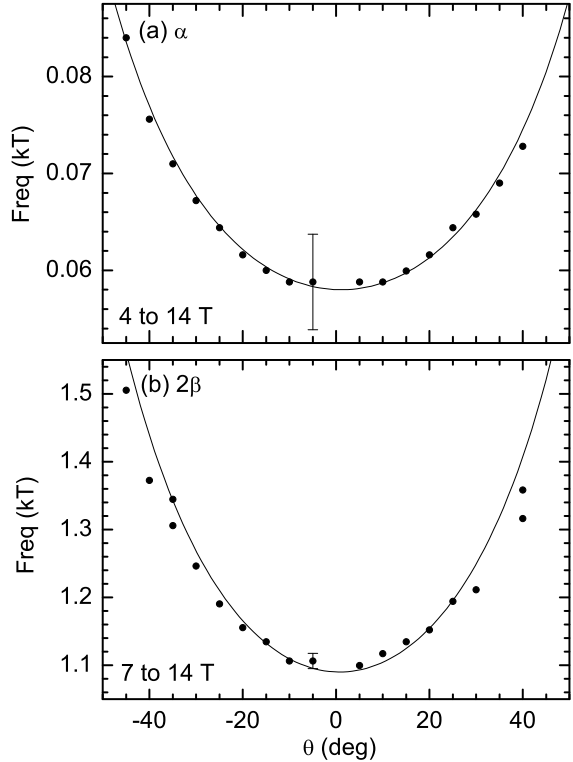}
\caption{\label{fig:angles}The (a)$\alpha$ and (b)2$\beta$ frequency peaks in the Fourier transform for the torque measurements of LaTe$_3$ at 2 K.  The field is oriented in the $b$-$c$ plane with an angle $\theta$ measured with respect to the $b$-axis. Solid curves show $1/\cos\theta$.  Error bars show FWHM for representative data points.}
\end{figure}

 Fig.~\ref{fig:angles} shows the results of angle-dependent measurements for the additional frequencies (a) $\alpha$ (FFT from 4 to 14 T) and (b) 2$\beta$ (FFT from 7 to 14 T). In each figure, a line is plotted representing $1/\cos\theta$.  As before, there is a close correspondence to $1/\cos\theta$ indicating that the FS sheets are indeed quasi-2D.  It is not surprising that the harmonic 2$\beta$ shows quasi-2D behavior, as its angle-dependence should be the same as that of its fundamental.  2$\beta$ splits into two at higher angles, but the peak is too weak to show the detailed structure that was seen in the Fourier transform of its fundamental, $\beta$.

\begin{figure}
\includegraphics[width=2.9in]{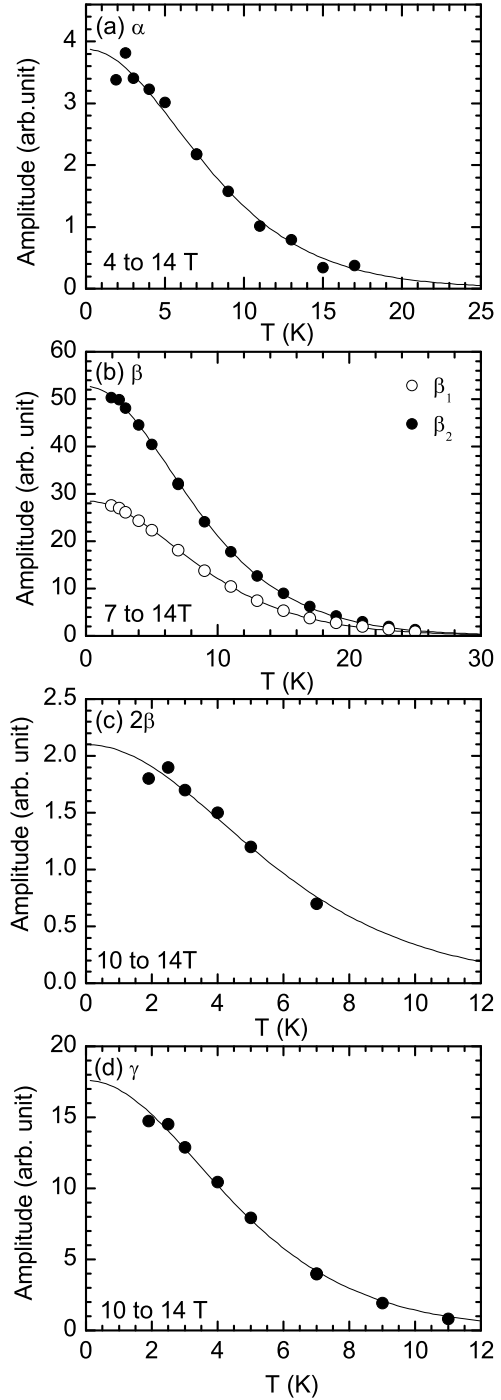}
\caption{\label{fig:temperature}Amplitudes as a function of temperature for LaTe$_3$, from torque measurements with field oriented 5$^\circ$ from the $b$-axis.  a)$\alpha$ (0.058 kT), (b)$\beta_1$ (0.52 kT) and $\beta_2$ (0.55 kT), (c)2$\beta$ (1.1 kT), and (d)$\gamma$ (1.6 kT).   Solid lines show fits to the Lifshitz-Kosevich relation, as described in the text.  The field range used for each set of Fourier transforms is indicated in each panel. }
\end{figure}

To obtain the effective mass $m^*$, torque measurements were extended from temperatures of 1.9 to 25 K, with the magnetic field offset from the crystallographic $b$-axis by 5$^\circ$.  The amplitudes of each frequency as derived from the Fourier transforms are shown in Fig.~\ref{fig:temperature} for (a)$\alpha$, (b)$\beta_1$ and $\beta_2$, (c)2$\beta$ and (d)$\gamma$. The field range used for each set of Fourier transforms is indicated in each panel.  The amplitudes decrease as temperature is increased.  Fits to the Lifshitz-Kosevich expression\cite{shoenberg} result in values for the cyclotron masses, which are shown in Table~\ref{frequencies}.  As expected, the effective mass associated with the 2$\beta$ frequency is twice that of $\beta$, confirming our assignation of this frequency as a higher harmonic.  Consequently, in Table~\ref{frequencies} we give the associated FS area as half that estimated from the Onsager relation.\cite{shoenberg}

\section{Discussion}

The two sets of measurements shown for fields oriented $H \| b$ (or close to $H \| b$ for the torque measurement) differ primarily in the number of split peaks observed in the $\beta$ frequencies, with up to six observed in the susceptibility measurement in Fig.~\ref{fig:dilution}, and two observed in the torque measurement in Fig.~\ref{fig:PPMS}.  The difference is unlikely to be due to the lower temperature of the susceptibility measurement as there are only minimal changes in the oscillation amplitudes for temperatures below 2 K, apparent from the Lifshitz-Kosevich curves in Fig.~\ref{fig:temperature}.  However, multiple peaks can arise if a sample contains multiple crystallites oriented at subtly different angles.  The fewer split peaks observed in the torque measurements, as well as the angle dependent measurements that show symmetry about $\theta$ = 0, imply that any influence from secondary misoriented crystallites in the torque sample is small.   While it is possible that the smaller field range used for the torque measurement could have the effect of ``merging'' closely spaced peaks, the well-formed nature of the double $\beta$ peaks in Fig.~\ref{fig:dilution}(b) implies that this may not be the case. In consequence, and for the sake of simplicity, further analysis of the $\beta$ frequency in this paper is
based on the torque results.

As indicated by ARPES\cite{brouet}, the small diamond-shaped pockets centered around the X point of the BZ (Fig.~\ref{fig:multifigure}) remain unmodified in the presence of the CDW.  Band structure calculations were performed for the unmodulated structure of LaTe$_3$, using the linear muffin-tin orbital (LMTO) method within the atomic sphere approximation, including combined-correction terms.\cite{barbiellini}  These calculations show that the inner X pocket has minimal and maximal FS areas corresponding to 2.14\% and 2.79\% of the BZ area, while the outer X pocket has minimal and maximal areas of 3.68\% and 3.82\% of the BZ area.\cite{Xpocket}  The measured $\beta$ frequencies, which correspond to similar FS areas, can thus be ascribed to these unmodified X pockets.

While in principle each of these four distinct frequencies should be observed in the measured data at $\theta$ = 0, in practice the finite Fourier transform resolution for the experimental conditions results in just two distinct frequencies at $\theta$ near zero.  As the LMTO band structure calculations indicate that the bilayer splitting is larger than the $b^*$-axis dispersion for both sheets (Fig.~\ref{fig:multifigure}(c)), $\beta_1$ and $\beta_2$ are tentatively assigned to the inner and outer X pockets respectively. An alternative interpretation would be that $\beta_1$ and $\beta_2$  are associated with the minimum and maximum of the inner X pocket, which would give a closer correspondence between calculated and measured areas. However this would imply that no frequencies are observed from the outer X pocket, even though both sheets are expected to be present in the reconstructed FS.

As $\theta$ is increased from zero, each of $\beta_1$ and $\beta_2$ splits into two peaks, and it is tempting to ascribe these split peaks $\beta_{1A}$ and $\beta_{1B}$ to the neck and belly frequencies of the inner X pocket, and similarly for $\beta_{2A}$ and $\beta_{2B}$ to the neck and belly frequencies of the outer X pocket.  But a simple sinusoidal $b$-axis dispersion should give neck and belly frequencies that are distinct at $\theta=0$ but then converge with increasing $\theta$ before meeting at a Yamaji angle\cite{bergemann_review, yamaji}.   The divergent behaviors of $\beta_{1A}$ and $\beta_{1B}$, and $\beta_{2A}$ and $\beta_{2B}$, could reflect deviations from a simple sinusoidal $b^*$-axis dispersion.  Alternatively, given that LaTe$_3$ is prone to twining due to the weak Van der Waals bonding between the double Te sheets, it is possible that there are grains in the sample that are misoriented within the plane by 90$^\circ$, so that as $\theta$ increases some grains could be rotating around the $a$-axis while other grains are rotating around the $c$-axis.  If these different rotation directions have different angle-dependencies, then this could cause single peaks near $\theta$ = 0 to split into multiple peaks at higher angles.

The measured cyclotron masses for the $\beta_1$ and $\beta_2$ frequencies are $m^*$ = 0.175(1)$m_0$ and 0.183(1)$m_0$, respectively.  These can be compared with the band masses $m_b$ for undressed quasiparticles where $m^*/m_b$ = 1 + $\lambda$ and $\lambda$ is the electron-phonon coupling parameter.  LMTO calculations show that $m_b$ = 0.113(5)$m_0$ for the inner X pocket, and $m_b$ = 0.154(8)$m_0$ for the outer X pocket.

The LMTO calculations from the average structure do not reveal any FS sheets with areas close to those given by the $\alpha$ and $\gamma$ frequencies.  The $\alpha$ and $\gamma$ frequencies are therefore ascribed to portions of the reconstructed FS.  The splitting of $\gamma$ from one peak to three peaks at higher angles, like the splitting of $\beta_1$ and $\beta_2$, can be attributed to either a complex $b^*$-axis dispersion or to crystal twinning.  Our measurements clearly show that the reconstructed FS is, as anticipated, quasi-2D. Inspection of the geometry of the FS of the unmodulated structure (Fig.~\ref{fig:multifigure}(c)) indicates that the precise geometry of the reconstructed FS would be very sensitive to details of the bilayer splitting.

\begin{figure}
\includegraphics[width =3.3 in]{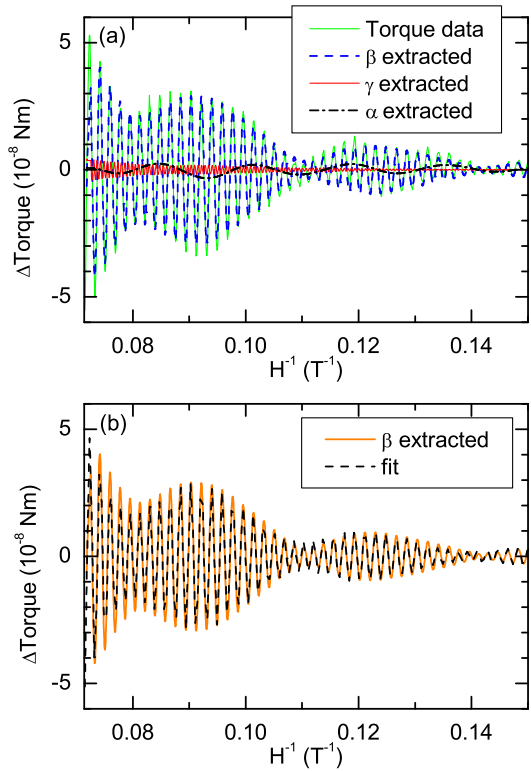}
\caption{\label{fig:extracted}(color online) Individual frequencies extracted from the filtered Fourier transform of a torque measurement of LaTe$_3$ at 2 K. (a) The extracted frequencies $\beta$ (blue, dashed), $\gamma$ (red, solid), and $\alpha$ (black, dash-dot) plotted against the raw data (green, solid).  (b) The extracted frequency $\beta$ oscillations (orange, solid) and a fit to the oscillations (black, dashed) as described in the text.}
\end{figure}

 We resort to an analysis of the field dependence of the quantum oscillations to provide further insight.  The torque data from Fig.~\ref{fig:PPMS} are shown again  in Fig.~\ref{fig:extracted}.  The frequencies $\alpha$, $\beta$ and $\gamma$ are individually extracted by filtering the Fourier transformed data with a gaussian filter.  The raw data is predominantly characterized by the frequency $\beta$ with its beating.  The extracted $\alpha$ and $\gamma$ frequencies show no beating, and their amplitude rises smoothly with an exponential dependence.

The mean-free path can be estimated by analyzing the exponential dependence of the oscillation amplitude with field. This exponent is given by $B_D = \frac{\hbar}{2e}\frac{C_F}{l}$ where $C_F$ is the circumference of the orbit in $k$-space and $l$ is the mean-free path.\cite{bergemann_review}  To solve for $B_D$, the extracted $\beta$ oscillations are fit to $A_0 e^{-B_D/H}(A_1$sin$(\beta_1/H+\phi_1) + A_2 $sin$(\beta_2/H+\phi_2))$ where $A_1$ and $A_2$ are the relative amplitudes of the $\beta_1$ and $\beta_2$ peaks in the Fourier transform, respectively and $A_0$ is a constant.  The factors $A_0$ and $B_D$ are allowed to vary alternately with the phases $\phi_1$ and $\phi_2$.  Fig.~\ref{fig:extracted}(b) shows the extracted $\beta$ oscillations along with the results of the fit.  The extracted oscillations are well-characterized by this expression.  From this fit, performed on several data sets for this sample, an exponent of $B_D$ = 37(1) T is obtained.  To find $C_F$, the constant of proportionality $R$ between the circumference and area, where $C_F = R * \sqrt{area}$, is derived from the calculated FS.  Using these values of $R$ ($R$ = 4.05 for the inner and 4.22 for the outer X pockets) and the measured areas for the $\beta_1$ and $\beta_2$ orbits, values for $C_F$ are obtained that result in a  mean free path of 82(4) nm.

Assuming that all sheets are characterized by the same mean free path, this value can in turn be used to estimate the $C_F$ and thus the shape for $\gamma$ orbit.  Fitting the extracted $\gamma$ oscillations to an exponential dependence results in $B_D$ = 52(6) T, and using the previously obtained value of $l$ leads to $C_F$ = 1.2(1) $\AA^{-1}$, with an $R$ value of 3.05.  This is smaller than the smallest possible value for $R$ (that of a circle with $R_{circle}$ = 2$\sqrt{\pi}$ = 3.54), which presumably reflects uncertainty in the exponents extracted from both the $\beta$ and $\gamma$ orbits.  Nevertheless, the implication that the $\gamma$ orbit is associated with a FS sheet that has a large cross-sectional area with respect to the circumference indicates that the geometry of this largest section of reconstructed FS is not significantly elongated (as implied by ARPES measurements)\cite{brouet_long}, but rather more closely approximates a circle.

Repeating the same analysis for the $\alpha$ frequency results in a circumference of 0.44(2) $\AA^{-1}$ which corresponds to an $R$ value of 5.8(3), which is significantly larger than that of a circle.  If the FS were an ellipse with the given area and circumference, it would have a semi-major axis 5 times greater than its semi-minor axis (0.097 $\AA^{-1}$ vs. 0.0185 $\AA^{-1}$).

Finally, we comment on the fraction of the FS volume in the CDW state relative to that of the unmodified FS. Optical conductivity measurements have suggested that for LaTe$_3$ as little as 2\% of the original FS volume remains in the gapped state.\cite{sacchetti}  Given that the $p_x$ and $p_z$ bands are approximately 5/8 filled in the unmodulated structure,\cite{yao} and that the 2D nature of the FS gives an easy correspondence between FS volume and area, this roughly corresponds to a FS area of 2\% of 5/8 of the BZ for the gapped state.  Our larger values of 7.65(5)\% of a BZ for $\gamma$, and 2.44(1)\% and 2.60(1)\% for $\beta_1$ and $\beta_2$, indicate that substantially more of the FS remains in the gapped state than is indicated by the optical conductivity results.\cite{safety_footnote}

\section{Conclusions}

De Haas van Alphen measurements on LaTe$_3$ show three fundamental frequencies.  A pair of frequencies $\beta_1$ and $\beta_2$ representing areas of 2.44(1)\% and 2.60(1)\% of the Brillouin zone, respectively, can be correlated with the unmodified X pocket of the FS from the average structure, a bilayer-split pocket which has been confirmed by ARPES to exist in the CDW state.  The other frequencies---$\alpha$, corresponding to 0.28(1)\% of the BZ and $\gamma$ corresponding to 7.65(5)\% of the BZ---must result from FS sheets reconstructed by the CDW wavevector.  These reconstructed FS sheets are quasi-2D, and analysis of the field dependence of the oscillations indicates that the geometry of the $\gamma$ sheet is approximately circular whereas that of the $\alpha$ sheet is elongated.

\section*{Acknowledgments}
We thank V. Brouet, L. Degiorgi, J.-H. Chu, J. Analytis for helpful discussions.  Scripts written by S. A. Grigera were used for data analysis.
This work is supported by the DOE, Office of Basic Energy Sciences, under contract number DE-AC02-76SF00515.


\begin{thebibliography}{10}
\bibitem{Xpocket}
Note that an alternative maximal area of 3.97\% might be expected from a
  slightly different connectivity of these Fermi surface sheets.
\bibitem{ellipse}
The approximation $2\pi\sqrt{(a^2 +b^2)/2}$ was used for the circumference of
  an ellipse.
\bibitem{safety_footnote}
Throughout the paper we have ignored the incommensurate nature of the gapping
  in discussing the dHvA oscillations. This incommensurability may be related
  to the presence of the extra frequencies marked with arrows in
  Fig.~\ref{fig:dilution}(c), but should not alter the basic conclusion that
  the FS volume in the CDW state is larger than that inferred from the optical
  conductivity.
\end{thebibliography}

 \end{document}